\documentclass[fleqn,usenatbib]{mnras}
\usepackage{newtxtext,newtxmath}
\usepackage[T1]{fontenc}

\DeclareRobustCommand{\VAN}[3]{#2}
\let\VANthebibliography\thebibliography
\def\thebibliography{\DeclareRobustCommand{\VAN}[3]{##3}\VANthebibliography}

\usepackage{graphicx}	
\usepackage{amsmath}	
\usepackage{float}
\usepackage{verbatim}
\usepackage[ ]{quoting} 
\usepackage{textcomp, gensymb}
\usepackage{multirow}
\usepackage{hyperref}
\usepackage{url}
\usepackage{amsmath}
\usepackage{enumitem}
\usepackage{subcaption}

\title[Carbon-deficient stars from APOGEE]{A large sample of newly-identified carbon-deficient red giants from APOGEE}

\author[Maben et al.]
{
Sunayana Maben,$^{1,2,3}$
Yerra Bharat Kumar,$^{3}$ \thanks{E-mail: bharat.yerra@iiap.res.in (YBK)}
Bacham E. Reddy,$^{3}$
Simon W. Campbell$^{4,5}$
and Gang Zhao$^{1,2}$ \thanks{E-mail: gzhao@nao.cas.cn (GZ)}
\\
$^{1}$CAS Key Laboratory of Optical Astronomy, National Astronomical Observatories, Chinese Academy of Sciences, Beijing 100101, Peoples Republic of China\\
$^{2}$School of Astronomy and Space Science, University of Chinese Academy of Sciences, Beijing 100049, Peoples Republic of China\\
$^{3}$Indian Institute of Astrophysics, II Block, Koramangala, Bangalore, 560034, India\\
$^{4}$School of Physics and Astronomy, Monash University, Clayton, Victoria, Australia\\
$^{5}$ARC Centre of Excellence for Astrophysics in Three Dimensions (ASTRO-3D), Australia}

\date{Accepted XXX. Received YYY; in original form ZZZ}

\pubyear{2023}

\begin{document}
\label{firstpage}
\pagerange{\pageref{firstpage}--\pageref{lastpage}}
\maketitle

\begin{abstract}
      Based on the APOGEE survey we conducted a search for carbon-deficient red giants (CDGs). We found 103 new CDGs, increasing the number in the literature by more than a factor of 3. CDGs are very rare, representing $0.03$~per cent of giants. They appear as an extended tail off the normal carbon distribution. We show that they are found in all components of the Galaxy, contrary to previous findings. The location of CDGs in the Hertzsprung-Russell diagram (HRD) shows that they are primarily intermediate-mass stars ($2-4~\rm{M}_{\odot}$). Their extended distribution may indicate that CDGs can also sometimes have  $M < 2.0~\rm{M}_{\odot}$. We attempted to identify the evolutionary phases of the CDGs using stellar model tracks. We found that the bulk of the CDGs are likely in the subgiant branch or red clump phase, whereas other CDGs may be in the red giant branch or early asymptotic giant branch phases. Degeneracy in the HRD makes exact identification difficult. We examined their C, N, and O compositions and confirmed previous studies showing that the envelope material has undergone extensive hydrogen burning through the CN(O) cycle. The new-CDGs have [C+N+O/Fe] that generally sum to zero, indicating that they started with scaled-solar composition. However, the previously known-CDGs generally have [C+N+O/Fe$] > 0.0$, indicating that some He-burning products were added to their envelopes. As to the site(s) in which this originally occurred, we do not find a convincing solution. 
 
\end{abstract}


\section{Introduction} \label{sec:intro}
    The weak G-band (wGb) giants are a rare class of chemically peculiar stars. More than a century ago \citet{Cannon1912} noticed the unusual spectrum of the red giant star HR~885 in which the G-band, the CH molecular absorption, was not well defined, unlike in spectra of other G- and K-type stars. \citet{Bidelman1951} examined the spectrum of HR~885 in detail and found that the G-band was absent. Since the star had other spectral features compatible with its spectral type (G5 III), he suggested that HR~885 ``may be a unique case of low carbon abundance". \citet{Bidelman1973} searched for more wGb stars among a sample of bright giants and found a total of 34 wGb giants based on how weak the CH G-band was.
    Subsequent detailed abundance analysis based on high-resolution spectra revealed that wGb stars are extremely carbon-deficient, nitrogen-enriched, and in some cases, overabundant in lithium \citep[see, e.g.][]{Sneden1978,Rao1978,Cottrell1978,Parthasarathy1980,Lambert1984,Parthasarathy1984}. Hereafter, we refer to wGb red giant stars as carbon-deficient giants (CDGs).

    After a long hiatus of about three decades, a renewed interest was found in decoding the puzzling nature of the peculiar chemical composition of the CDGs. Recent studies analysed the high-resolution spectra of the known-CDGs \citep[$R \approx 48,000 -  60,000$;][] {Adamczak2013,Palacios2016} with modern tools incorporating the latest stellar model atmospheres, and  atomic and molecular data for accurate stellar parameters, abundances, and isotopic ratios. Results from these studies are in good agreement with each other -- the carbon abundances of CDGs are under-abundant by about a factor of 20 when compared to normal giants, the carbon isotopic ratios $\rm ^{12}C/^{13}C$ are close to the equilibrium value of $3 - 4$, and the nitrogen abundances are highly enhanced. In some cases, the CDGs are also found to be overabundant in Li and Na. Studies also reached a consensus on their masses: the CDGs are of intermediate-mass ranging from $2.5 - 5~\rm{M}_{\odot}$. Currently, there is no consensus on any particular scenario which explains the chemical composition of the CDGs. Some favour in-situ origin, that is, internal nucleosynthesis and non-standard mixing in stars \citep{Adamczak2013}, and others favour  external origin such as pollution of stellar atmospheres during stars' main sequence or pre-main sequence evolution by the CN-processed material \citep{Palacios2016}. 

    To date, there are not many wGb giants known, just 44 in the literature, according to the latest compilation (\citealt{Bidelman1951}; \citealt{Bidelman1973}; \citealt{Bond2019}). Increasing the sample size may help identify the evolutionary link with the carbon deficiency, identifying their progenitors or descendants. The work by \citet{Bidelman1973} laid a foundation for finding more of these peculiar stars. \cite{Bond2019} recently added 5 CDGs based on spectroscopy to the initial list of CDGs. Since then, no new CDGs based on spectroscopy have been identified. Recent studies by \citet{Adamczak2013}, \citet{Palacios2016}, and especially \citet{Bond2019} highlighted the need for a larger number of CDGs to understand their origin.

    With the advent of high-resolution spectroscopic surveys, it is now possible to systematically search for carbon-deficient giants among a large sample of giants. The current study aims to significantly increase the sample of CDGs by taking advantage of these new surveys. We also include an initial analysis of the new CDG sample. Our next paper provides a deeper analysis by including asteroseismic data. This is the largest systematic search for CDGs since the initial spectroscopic survey among bright stars by \citet{Bidelman1973}.

    \begin{figure*}
    \centering
    \hspace*{-0.5cm}\includegraphics[width=\linewidth]{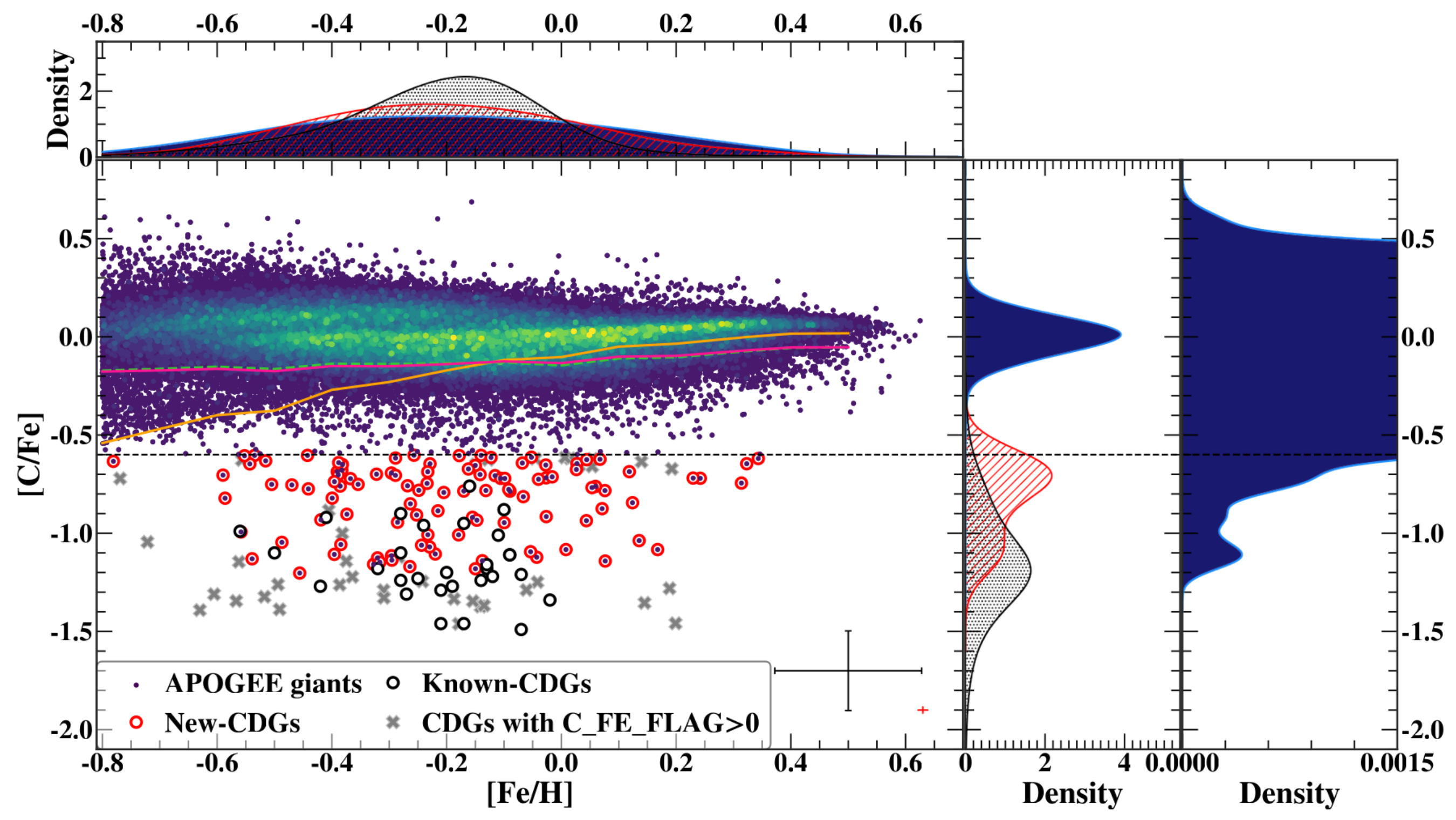}
    
    \caption{Carbon abundances against metallicity for the selected sample giants from the APOGEE~DR17 survey (APOGEE giants). The colour scale represents the number density of stars; the lighter greenish colour indicates a higher density. Also shown are the newly identified CDGs (new-CDGs; open red circles) and the known-CDGs from the literature (open black circles). As the C abundances of the known-CDGs are from optical spectra, we offset their C abundances by $+0.02$~dex to match the APOGEE abundance scale (see Section~\ref{sec:cno_results}). The APOGEE CDG sample having possible issues (\texttt{C\_FE\_FLAG}~>~0) associated with carbon abundance measurements (filled grey crosses) is also shown. Kernel density histograms are included to show the [Fe/H] and [C/Fe] distributions of the different samples, with the colour scheme following the background sample. Average error bars are shown on the bottom right end of the left panel. The average uncertainties of the carbon abundances are used for the individual Gaussian widths of the KDE. We find a broad carbon distribution for the known-CDGs. The observed [C/Fe] values are compared with the predicted [C/Fe] ratios at the RGB tip from standard models for initial masses 1~M$_{\odot}$ (orange solid line), 2~M$_{\odot}$ (green dashed line) and 3~M$_{\odot}$ (pink solid line). The horizontal dashed line at [C/Fe]~$= -0.6$~dex denotes the cut we use to identify carbon-deficient stars (see Section~\ref{sec:CDG_identification}). The adopted solar abundances for all stars are from \citet{Grevesse2007}. See text for details.}
    \label{fig:fig1}
    \end{figure*}

\section{Sample Selection} \label{sec:data}
    We used  sample stars from data release 17 (DR17; \citealt{Abdurrouf2022}) of the Apache Point Observatory Galactic Evolution Experiment (APOGEE; \citealt{Majewski2017}), which is a large-scale near-IR (H-band), high-resolution ($R \approx 22,500$) spectroscopic survey. DR17 is the final data release and contains data from the northern and southern hemispheres. The survey uses twin high-resolution spectrographs \citep{Wilson2019} mounted on Apache Point Observatory’s  $2.5-$m  Sloan  Telescope \citep{Gunn2006} and  Las  Campanas Observatory’s $2.5-$m du Pont telescope \citep{Bowen1973}. Data was reduced using automatic reduction pipelines \citep{Nidever2015}. The data release contains derived radial velocities, stellar parameters ($T_{\rm eff}$, log $g$), and elemental abundances of 20 species from C to Ce for 733,901 stars across the Milky Way (APOGEE Stellar Parameters and Chemical Abundance Pipeline, ASPCAP; \citealt{Garcia2016}).

    Intending to increase the probability of detecting carbon-deficient giants, we restricted the sample to stellar parameters encompassing the known-CDGs in the literature. We restricted our sample to disk metallicities ([Fe/H]~$\geq -0.8$~dex; \citealt{Tomkin1995}) to be consistent with that of the known-CDGs. The known-CDGs are evolved off the main sequence; hence we restricted the sample to surface gravity, log~$g$~$\leq 3.8$~dex. The catalogue had many stars from LMC and SMC, which were removed using the criterion given in \cite{Hasselquist2021}. Further, we culled all the giants with at least one uncertain parameter flagged in the catalogue as \texttt{STAR\_BAD} or \texttt{STAR\_WARN}. `\texttt{STAR\_BAD}' is set if any of \texttt{TEFF} (effective temperature), \texttt{LOGG} (surface gravity), \texttt{CHI2} (poor matches to synthetic spectra), \texttt{COLORTE} (photometric temperature), \texttt{ROTATION} (rotation/broad lines), \texttt{SN} (signal-to-noise) is bad or any individual parameter is near grid edge. The \texttt{STAR\_WARN} is set if any of \texttt{TEFF}, \texttt{LOGG}, \texttt{CHI2}, \texttt{COLORTE}, \texttt{ROTATION}, \texttt{SN} is uncertain. The carbon abundances in the catalog (\texttt{C\_FE}) are determined from CO and CN molecular lines (primarily CO lines; $15000.455$~\AA, $15508.429$~\AA, $16071.736$~\AA, $16536.181$~\AA, $17040.075$~\AA; \citealt{Li2015}). We picked giants only bearing \texttt{C\_FE\_FLAG} = 0 for high-quality carbon-to-iron abundance ratio measurements. The abundances in the APOGEE catalog are flagged\footnote{\url{https://www.sdss4.org/dr17/irspec/abundances/}} (\texttt{X\_FE\_FLAG}), with zero being the most reliable \citep{Jonsson2020}. With these constraints, we obtained a final sample of 315,789 giants with quality abundances (hereafter, `APOGEE giants').

\section{Analysis and Results} \label{sec:results}

\subsection{[C/Fe] Distribution \& CDG identification} \label{sec:CDG_identification}
    The final sample from APOGEE that made it through the quality criteria discussed in Section~\ref{sec:data} is shown in the [C/Fe]-[Fe/H] plane in Fig.~\ref{fig:fig1}. We show the kernel density estimation (KDE) distribution of [C/Fe] in the right panel of the figure. The bulk of the giants falls in a main peak at [C/Fe$] = 0.0$~dex with 1$\sigma$ dispersion of $0.1$~dex. The distribution also shows a tail to very low [C/Fe].

    To check the theoretically expected range of [C/Fe] in giants, we used stellar models from the MIST database (MESA isochrones and Stellar Tracks; \citealt{Dotter2016,Choi2016}). In Fig.~\ref{fig:fig1}, we plot lines representing the carbon abundances at the tip of the red giant branch (RGB) for giants of three different masses, 1, 2, and 3~M$_{\odot}$, for a range of metallicity values. It can be seen that the intermediate-mass models, with shallow convective envelopes, show only a modest depletion of carbon (due to the first dredge-up) with decreasing [Fe/H] reaching [C/Fe$] \approx -0.2$~dex. However, in low-mass models (e.g. 1~M$_{\odot}$), carbon depletion increases with decreasing metallicity, reaching as low as [C/Fe$] \approx -0.6$~dex at [Fe/H$]= -0.8$~dex. Considering this, we adopt a conservative limit of [C/Fe$] = -0.6$~dex as the defining line for CDGs. That is,  all giants with [C/Fe$] < -0.6$~dex are classified as CDGs. We note that the limit is $6 \sigma$ away from the peak of the main [C/Fe] component (Fig.~\ref{fig:fig1}). Between the values of [C/Fe$] = -0.6$~dex and $-0.1$~dex, there are still several giants, particularly at the metal-rich regime, which are mildly deficient in carbon compared to model predictions.  Many of these probably fall under the category of CDGs. But, in this study, we have not considered them as CDGs to avoid  possible contamination with normal giants from the main distribution.

    With the adopted criteria, we identify 103 new-CDGs\footnote{The weak CH band in the spectra of one of our new-CDGs, with APOGEE ID 2M16405512+6435204, was previously noted by \cite{Keenan&McNeil1989}. We thank the referee for pointing this out. We discuss the \cite{Keenan&McNeil1989} study in Section~\ref{sec:CDG_identification}.}. This finding  increases the CDG sample to more than a factor of three from the  44 known-CDGs in the literature (\citealt{Cannon1912}; \citealt{Bidelman1951}; \citealt{Roman1952}; \citealt{Bidelman1957}; \citealt{Greenstein1958}; \citealt{Spinrad1969}; \citealt{Bidelman1973}; \citealt{Bond2019}). Of the known-CDGs, only 29 have carbon abundances from high-resolution optical spectra \citep{Adamczak2013,Palacios2016}, so our new sample represents an increase by a factor of 4.6 in this sense. The new-CDGs are listed in Table~\ref{tab:table1}.

\begin{table*}
\centering
\caption{Basic and atmospheric parameters along with the known and new carbon-deficient giants' CNO abundance ratios. Note that offsets of $+0.02$$\sim$dex, $-0.15$$\sim$dex, $-0.12$$\sim$dex, and $-0.03$$\sim$dex have been applied to the {[}C/Fe{]}, {[}N/Fe{]}, {[}O/Fe{]}, and {[}Fe/H{]} abundance ratios of the known carbon-deficient giants to put them on the same abundance scale as APOGEE (see Section~\ref{sec:cno_results}). The full sample will be provided online and in the supplementary online-only material.}
\label{tab:table1}
\resizebox{\linewidth}{!}{%
\begin{tabular}{lrrrccccccccc}
\hline
\multicolumn{1}{c}{Star} & \multicolumn{1}{c}{R.A.}      & \multicolumn{1}{c}{Dec.} & H$^{\ddag}$     & V     & E(B-V) & $\log (L/L_{\odot})$ & $T_{\rm eff}$ & {[}Fe/H{]} & log $g$       & {[}C/FE{]} & {[}N/FE{]} & {[}O/FE{]} \\
                    & \multicolumn{1}{c}{(J2000)}     & \multicolumn{1}{c}{(J2000)}      &       &       &        &                      & (K)           &            &  &            &            &            \\
\hline
\hline
HD 40402$^{**}$     & 89.32889  & $-27.33151$     &   6.47    & 8.61  & 0.05   & 2.16$\pm$0.05        & 5005          & $-0.17$      & 2.80           & $-0.95$      & $+1.08$       & $+0.20$        \\
HD 124721$^{**}$    & 214.10232 & $-45.18977$      &    7.19   & 9.54  & 0.11   & 2.13$\pm$0.04        & 5107          & $-0.07$      & 2.64          & $-1.21$      & $+1.05$       & $+0.18$       \\
37 Com$^{**}$    & 195.06861 & $+30.78502$        &   2.38    & 4.84  & 0.00      & 2.66$\pm$0.05        & 4610          & $-0.56$      & 2.50           & $-0.99$      & $+2.40$        & $+0.50$        \\
HR 885           & 44.95745  & $+47.22071$         &  3.19     & 5.47  & 0.00      & 2.15$\pm$0.04        & 5198          & $-0.17$      & 2.65          & $-1.46$      & $+1.03$       & $+0.14$       \\
HR 1023$^{**}$   & 50.91246  & $+4.88210$        &    4.57   & 6.42  & 0.12   & 2.39$\pm$0.06        & 5310          & $-0.25$      & 1.60           & $-1.23$      & $+1.69$       & $+0.08$       \\
HR 1299        & 62.69113  & $-35.27379$           &     4.23  & 6.44  & 0.01   & 2.38$\pm$0.04        & 4690          & $-0.11$      & 2.20           & $-1.01$      & $+1.30$        & $-0.05$      \\
HD 28932         & 68.38199  & $+1.02247$         &     5.57  & 7.95  & 0.08   & 2.30$\pm$0.04         & 4915          & $-0.42$      & 2.50           & $-1.27$      & $+1.36$       & $+0.25$       \\
HD 31869     & 74.54692  & $-28.03567$            &    7.11   & 9.29  & 0.03   & 2.30$\pm$0.05         & 4800          & $-0.50$       & 1.80           & $-1.10$       & $+1.26$       & $+0.19$       \\
HD 49960$^{**}$   & 102.41893 & $-31.25912$       &    5.99   & 8.35  & 0.11   & 2.28$\pm$0.06        & 5030          & $-0.20$       & 2.61          & $-1.20$       & $+1.18$       & $+0.32$       \\
HD 67728$^{**}$ & 122.15687 & $-19.83991$         &   5.15    & 7.54  & 0.01   & 2.82$\pm$0.05        & 4827          & $-0.24$      & 2.28          & $-0.96$      & $+1.33$       & $+0.18$       \\
\multicolumn{1}{c}{...} & ...       & ...  & ...   & ...   & ...    & ...                  & ...           & ...        & ...           & ...        & ...        & ...        \\

2M04403830+2554274     & 70.15962  & $+25.90763$    & 9.48  & 16.22 & 1.69   & 1.78$\pm$0.06        & 4992          & $+0.06$       & 2.41          & $-0.76$      & $+0.70$        & $-0.16$      \\
2M12261419-6325497    & 186.55916 & $-63.43049$    & 9.81  & 15.52 & 3.45   & 4.66$\pm$0.10         & 4667          & $+0.12$       & 2.01          & $-0.69$      & $+0.70$        & $-0.09$      \\
2M20244529+3712196   & 306.18875 & $+37.20547$     & 12.19 & 22.15 & 1.75   & 0.06$\pm$0.33        & 4908          & $-0.35$      & 2.11          & $-0.75$      & $+0.45$       & $-0.11$      \\
2M05481354+2926231    & 87.05644  & $+29.43978$    & 10.24 & 13.88 & 0.59   & 2.19$\pm$0.07        & 4720          & $-0.07$      & 2.12          & $-0.81$      & $+0.70$        & $+0.06$       \\
2M00344095+5804544    & 8.67067   & $+58.08179$    & 12.01 & 14.34 & 0.37   & 2.00$\pm$0.09           & 5089          & $-0.23$      & 2.82          & $-0.65$      & $+0.62$       & $+0.04$       \\
2M12254845-6228451     & 186.4519  & $-62.47921$   & 10.31 & 15.14 & 4.22   & 2.73$\pm$0.06        & 4820          & $+0.31$       & 2.43          & $-0.74$      & $+0.76$       & $-0.13$      \\
2M01450782+6445416  & 26.28262  & $+64.76156$      & 9.47  & 14.95 & 0.88   & 2.29$\pm$0.07        & 4392          & $-0.14$      & 1.63          & $-1.14$      & $+0.51$       & $+0.06$       \\
2M19055092+3745351    & 286.46221 & $+37.75975$    & 9.02  & 11.28 & 0.10    & 2.17$\pm$0.05        & 5007          & $-0.39$      & 2.28          & $-0.70$       & $+0.63$       & $+0.08$       \\
2M04222073+4844377  & 65.58641  & $+48.74382$     & 12.01 & 17.43 & 1.19   & 1.16$\pm$0.16        & 5137          & $-0.09$      & 2.60           & $-0.79$      & $+0.66$       & $-0.06$      \\
2M16405512+6435204$^{\dagger}$     & 250.22971 & $+64.58903$     & 2.33  & 4.84  & 0.02   & 2.83$\pm$0.04        & 4591          & $-0.10$       & 2.02          & $-0.72$      & $+0.51$       & $-0.05$      \\
\multicolumn{1}{c}{...} & ...       & ...  & ...   & ...   & ...    & ...                  & ...           & ...        & ...           & ...        & ...        & ...        \\
\hline
\end{tabular}%
}
\begin{quoting}
      { \textbf{Notes.}\\
      $^{(**)}$  Binary/ Multiple system.\\
      $^{(\ddag)}$ The 2MASS catalog \citep{Cutri2003}\\ 
      All the stellar designations beginning with `2M' denote the APOGEE ID of the new carbon-deficient giants.\\
      $^{(\dagger)}$ This star has been previously identified as a weak CH star by \cite{Keenan&McNeil1989}.}
\end{quoting}
\end{table*}

     In Fig.~\ref{fig:fig1}, we highlight the new-CDGs as red circles and also show the known-CDGs (that have reported [C/Fe] abundances) for comparison. The C abundances of the known-CDGs are from optical and the new-CDGs from IR spectra. We offset the C abundances of the known-CDGs by $+0.02$~dex to match the APOGEE abundance scale (see Section~\ref{sec:cno_results}). It can be seen that there is substantial overlap between the known- and new-CDGs: $50$~per cent of the new-CDGs have [C/Fe$] \leq -0.76$~dex, like the known-CDGs. The known-CDGs appear to have a more extreme carbon deficiency than the new-CDGs, on average. The known-CDGs have an average [C/Fe$] = -1.2$~dex compared to new-CDGs with an average [C/Fe$] = -0.8$~dex.

      Looking at the [Fe/H] distribution in the top panel of Fig.~\ref{fig:fig1}, it can be seen that the distributions of both the new- and known-CDGs peak in roughly at the same [Fe/H] as the background sample ($\approx -0.2$~dex). That the known-CDGs appear to have normal metallicity is at odds with the conclusion of \cite{Adamczak2013}, who reported that they are of systematically lower metallicity than the field (the background sample in \cite{Adamczak2013} was relatively small). The known-CDGs have a more peaked distribution, which may be due to the small sample size.

   We cross-matched our new-CDG sample with the study of \cite{Keenan&McNeil1989} and found one overlapping star. This study seems to have been  overlooked in the literature. The study doesn't focus on CDGs, but it does give rough classifications of CH band strength\footnote{Giants are classified as CH$-0.5$, CH$-1$, ...,  CH$-5$. Higher the index, greater the carbon deficiency.}. The one common star between the two studies, 18~Dra, (APOGEE ID 2M16405512+6435204) has a CH$-2$ classification in \cite{Keenan&McNeil1989}. Among the known-CDG sample, we found 13 CDGs in common with the \cite{Keenan&McNeil1989} study. Except for three of these stars, all were categorised by \cite{Keenan&McNeil1989} as `CH$-2$' or weaker.
   
  Further, we cross-matched the entire APOGEE catalog with the \cite{Keenan&McNeil1989} sample. We found six stars in common with APOGEE. The APOGEE [C/Fe] abundances show that only one of these stars, the star 18~Dra identified above, is a CDG by our definition. The other 5 stars are not CDGs and have CH classifications higher than CH$-2$. This means that the CH$-2$ (and weaker) stars of \cite{Keenan&McNeil1989} are generally more carbon deficient. This highlights the need to obtain high-resolution spectra to confirm carbon deficiency.

   As mentioned in Section~\ref{sec:data}, our new-CDG sample is based on high-quality data. We specifically limit the sample based on (i) quality stellar parameters (\texttt{STAR\_BAD} or \texttt{STAR\_WARN} should not be set), and (ii) high-quality [C/Fe] measurements (\texttt{C\_FE\_FLAG} = 0). The first cut restricts the sample to good-quality data and does not bias the [C/Fe] distribution; however, the second cut could have excluded some stars with very low (or very high) abundances, close to the outer range of models used by APOGEE ($-1.5 \leq [$C/Fe$] \leq +1.0$~dex)\footnote{\url{https://www.sdss4.org/dr17/irspec/apogee-libraries/}}. We note that ASPCAP does not provide abundance limits. To determine what proportion of CDGs were excluded, we removed the [C/Fe] quality constraints. With this criteria, we identify 39 more CDGs. In Fig.~\ref{fig:fig1} we show this CDG sample with filled grey crosses. It can be seen that a large number of CDGs that exhibit extreme deficiency in carbon just like the known-CDGs were excluded due to non-optimal [C/Fe] measurements. The other reason that many of these CDGs did not have their \texttt{C\_FE\_FLAG} = 0 is that there were issues identified during the spectral fitting process, and/or issues that occured when evaluating the stellar parameters\footnote{\url{https://www.sdss4.org/dr17/irspec/apogee-bitmasks/}}. These problems are flagged in the catalog as \texttt{VMICRO\_WARN}, \texttt{VSINI\_WARN}, \texttt{VSINI\_BAD}, and so on (see Table~\ref{tab:table2}). Thus our high-quality sample misses many stars with extreme carbon-deficiency. This highlights the need to obtain high-resolution spectra to determine good-quality carbon abundances and study these stars in detail. To facilitate follow-up studies, we list these 39 stars in Table~\ref{tab:table2}, along with their atmospheric parameters, [C/Fe] abundance ratios, \texttt{ASPCAPFLAGS}, and \texttt{C\_FE\_FLAG}. To be consistent with Table~\ref{tab:table1}, we report V magnitudes which were calculated from \textit{Gaia} G magnitudes using the colour-colour transformations given in \citet{Riello2021}. From here on we do not consider these stars further, apart from some discussion in Section~\ref{sec:cno_results}.

  Next, we investigate the kinematic properties, location in the Hertzsprung–Russell diagram (HRD), and CNO abundances of our new-CDGs and the known-CDGs. This may help in understanding the two groups.

\begin{table*}
\centering
\caption{The list of carbon-deficient giants whose \texttt{C\_FE\_FLAG}~$>~0$. The star identifications assigned by  APOGEE~DR17, the V magnitudes, the [C/Fe] abundance ratios along with the \texttt{C\_FE\_FLAG} and  \texttt{ASPCAPFLAGS} are provided. The bitmasks in the \texttt{C\_FE\_FLAG} are used to provide information about the possible issues associated with [C/Fe] abundance measurements. The \texttt{ASPCAPFLAGS} provides information and indicate possible issues associated with the ASPCAP fits. The table will be provided online and in the supplementary online-only material.}
\label{tab:table2}
\resizebox{\linewidth}{!}{%
\begin{tabular}{lrrrl}
\hline
\multicolumn{1}{c}{APOGEE~ID}          & \multicolumn{1}{c}{V}     & {[}C/FE{]} & \multicolumn{1}{c}{\texttt{C\_FE\_FLAG}} & \multicolumn{1}{c}{\texttt{ASPCAPFLAGS}}                                               \\
\hline
\hline
2M01285376+6420303 & 16.26 & $-1.46$      & 288             & \texttt{C\_M\_WARN}                                                \\
2M17023823-4204106 & 22.19 & $-1.46$      & 256             & \texttt{C\_M\_WARN}                                                \\
2M00423496-7324474 & 15.38 & $-1.39$      & 256             & \texttt{VMICRO\_WARN}, \texttt{C\_M\_WARN}                                   \\
2M05495914-6717120 & 16.13 & $-1.39$      & 288             & \texttt{C\_M\_WARN}                                                \\
2M05515652+2149291 & 15.23 & $-1.37$      & 256             & \texttt{C\_M\_WARN}                                                \\
2M18534451-3042206 & 15.14 & $-1.37$      & 256             & \texttt{C\_M\_WARN}                                                \\
2M22510184+5236454 & 13.46 & $-1.35$      & 256             & \texttt{C\_M\_WARN}                                                \\
2M07260470+4015017 & 10.59 & $-1.35$      & 256             & \texttt{C\_M\_WARN}                                                \\
2M05184812-6846111 & 15.42 & $-1.34$      & 256             & \texttt{VMICRO\_WARN}, \texttt{C\_M\_WARN}                                   \\
2M06264830+5505244 & 10.93 & $-1.33$      & 256             & \texttt{C\_M\_WARN}                                                \\
2M08011220+2618067 & 10.50  & $-1.33$      & 288             & \texttt{C\_M\_WARN}                                                \\
2M07095586+0632506 & 12.36 & $-1.32$      & 256             & \texttt{C\_M\_WARN}                                                \\
2M05542936-7047480 & 16.13 & $-1.31$      & 256             & \texttt{C\_M\_WARN}                                                \\
2M07240291-2552332 & 19.34 & $-1.29$      & 288             & \texttt{C\_M\_WARN}                                                \\
2M07271834-6700515 & 9.93  & $-1.29$      & 256             & \texttt{C\_M\_WARN}                                                \\
2M07375910-6529050 & 10.73 & $-1.28$      & 256             & \texttt{C\_M\_WARN}                                                \\
2M06113067+2808110 & 14.03 & $-1.26$      & 256             & \texttt{C\_M\_WARN}                                                \\
2M23533720-7314184 & 16.62 & $-1.26$      & 256             & \texttt{VMICRO\_WARN}, \texttt{C\_M\_WARN}                                   \\
2M05494354+3438113 & 12.51 & $-1.25$      & 256             &                                                           \\
2M08120643-3320409 & 15.26 & $-1.25$      & 256             &                                                           \\
2M06301992-1604284 & 14.45 & $-1.22$      & 32              &                                                           \\
2M06490254+2008508 & 13.67 & $-1.15$      & 32              &                                                           \\
2M04515993-6853269 & 14.91 & $-1.14$      & 32              & \texttt{VMICRO\_WARN}, \texttt{VMICRO\_BAD}                                  \\
2M03391361+6717213 & 15.18 & $-1.12$      & 32              &                                                           \\
2M07402109+2135570 & 13.40  & $-1.04$      & 32              &                                                           \\
2M04155246+2429467 & 14.23 & $-1.00$         & 32              &                                                           \\
2M07293026-1913194 & 16.80  & $-0.88$      & 32              &                                                           \\
2M07012477-7344326 & 18.04 & $-0.72$      & 32              &                                                           \\
2M00085560+7542082 & 10.61 & $-0.67$      & 256             & \texttt{VMICRO\_WARN}, \texttt{C\_M\_WARN}, \texttt{N\_M\_WARN}, \texttt{VSINI\_WARN}            \\
2M19485275+6448576 & 9.55  & $-0.66$      & 256             & \texttt{C\_M\_WARN}, \texttt{VSINI\_WARN}, \texttt{VSINI\_BAD}                         \\
2M20322594+4114054 & 18.27 & $-0.66$      & 288             & \texttt{VMICRO\_WARN}, \texttt{N\_M\_WARN}, \texttt{VSINI\_WARN}, \texttt{C\_M\_BAD}             \\
2M08290857+5340329 & 10.81 & $-0.65$      & 256             & \texttt{VMICRO\_WARN}, \texttt{C\_M\_WARN}, \texttt{N\_M\_WARN}, \texttt{VSINI\_WARN}, \texttt{VSINI\_BAD} \\
2M18311112-1153249 & 9.02  & $-0.64$      & 256             & \texttt{C\_M\_WARN}, \texttt{N\_M\_WARN}, \texttt{VSINI\_WARN}                         \\
2M07191122+5632490 & 13.49 & $-0.63$      & 288             & \texttt{VMICRO\_WARN}, \texttt{C\_M\_WARN}, \texttt{N\_M\_WARN}, \texttt{VSINI\_WARN}            \\
2M11012939+4656335 & 11.46 & $-0.63$      & 256             & \texttt{VMICRO\_WARN}, \texttt{C\_M\_WARN}, \texttt{VSINI\_WARN}                       \\
2M07555245+1945265 & 9.34  & $-0.62$      & 256             & \texttt{C\_M\_WARN}, \texttt{N\_M\_WARN}, \texttt{VSINI\_WARN}, \texttt{VSINI\_BAD}              \\
2M08070343+4155577 & 11.68 & $-0.62$      & 256             & \texttt{C\_M\_WARN}, \texttt{N\_M\_WARN}, \texttt{VSINI\_WARN}                         \\
2M10353072+4154353 & 9.12  & $-0.62$      & 256             & \texttt{VMICRO\_WARN}, \texttt{C\_M\_WARN}, \texttt{N\_M\_WARN}                        \\
2M19321881+1737581 & 9.96  & $-0.61$      & 256             & \texttt{VMICRO\_WARN}, \texttt{C\_M\_WARN}, \texttt{N\_M\_WARN}              \\
\hline
\end{tabular}%
}
\end{table*}

    \subsection{Kinematics and Galactic component membership} \label{sec:kinematics}

       Here we compile the kinematic properties of the known- and new-CDGs and determine what component(s) of the Galaxy they belong to. To compute the spatial distribution of the CDGs, we converted the ICRS coordinates ($\alpha$, $\delta$) into Galactic coordinates ($l$, $b$) using the \texttt{astropy.coordinates} package \citep{Astropy2013, Astropy2018}. We then transformed the Galactic coordinates and heliocentric distance for the CDGs into a Cartesian Galactocentric coordinate system (X, Y, Z)\footnote{The position component X points from the position of the Sun projected to the Galactic midplane to the Galactic center – roughly towards ($l$, $b$)~=~(0$^{\circ}$,0$^{\circ}$). The Y component points roughly towards the Galactic longitude $l$~=~90$^ {\circ}$, and the Z component points roughly towards the North Galactic Pole ($b$~=~90$^ {\circ}$).} using the Astropy \texttt{Galactocentric} package \citep{Astropy2013, Astropy2018}, and derived the Galactocentric distance (R)\footnote{R~=~($X^{2}$+$Y^{2}$+$Z^{2}$)$^{1/2}$}. We assumed that the Sun's distance from the Galactic center $R_{\odot}$~=~8.2~kpc \citep{Bland-Hawthorn2016}, and its distance above the Galactic plane $z_{\odot}$~=~25~pc \citep{Juric2008}. The distance from each star to the Sun was taken from the catalog of \citet{Bailer-Jones2021}.
       
         We calculated the Galactic space velocity components (U, V, W) for both the new- and known-CDG samples using astrometry with distances taken from the catalog of \citet{Bailer-Jones2021} and proper motions from \textit{Gaia}~DR3 \citep{Gaia2016, Gaia2022}. The radial velocity data were taken from APOGEE~DR17 for the new-CDGs and \textit{Gaia}~DR2 for the known-CDGs. The space velocity components were computed using the Astropy \texttt{Galactocentric} package \citep{Astropy2013, Astropy2018}. Space velocities  are converted to the LSR frame using the solar motion  ($U_{\odot}$, $V_{\odot}$, $W_{\odot}$)~= (11.1, 12.24, 7.25)~\,km\,s$^{-1}$ \citep{Bland-Hawthorn2016}. The velocity component U is taken to be positive in the direction toward the Galactic center, the V component is positive in the direction of Galactic rotation, and the W component is positive toward the North Galactic Pole. We adopt a LSR velocity $V_{LSR}$ = 232.8~\,km\,s$^{-1}$. The space velocities ($U_{LSR}$, $V_{LSR}$, $W_{LSR}$), along with radial velocities for all the stars\footnote{The new-CDG with \texttt{APOGEE\_ID} 2M20402480+4234303 did not have its proper motion and distance in the above-mentioned catalogs and hence its spatial distribution could not be estimated.}, are given in Table~\ref{tab:table3}.

        \begin{figure*}
        \centering
        \begin{subfigure}[b]{0.7\linewidth}
        \includegraphics[width=1\linewidth]{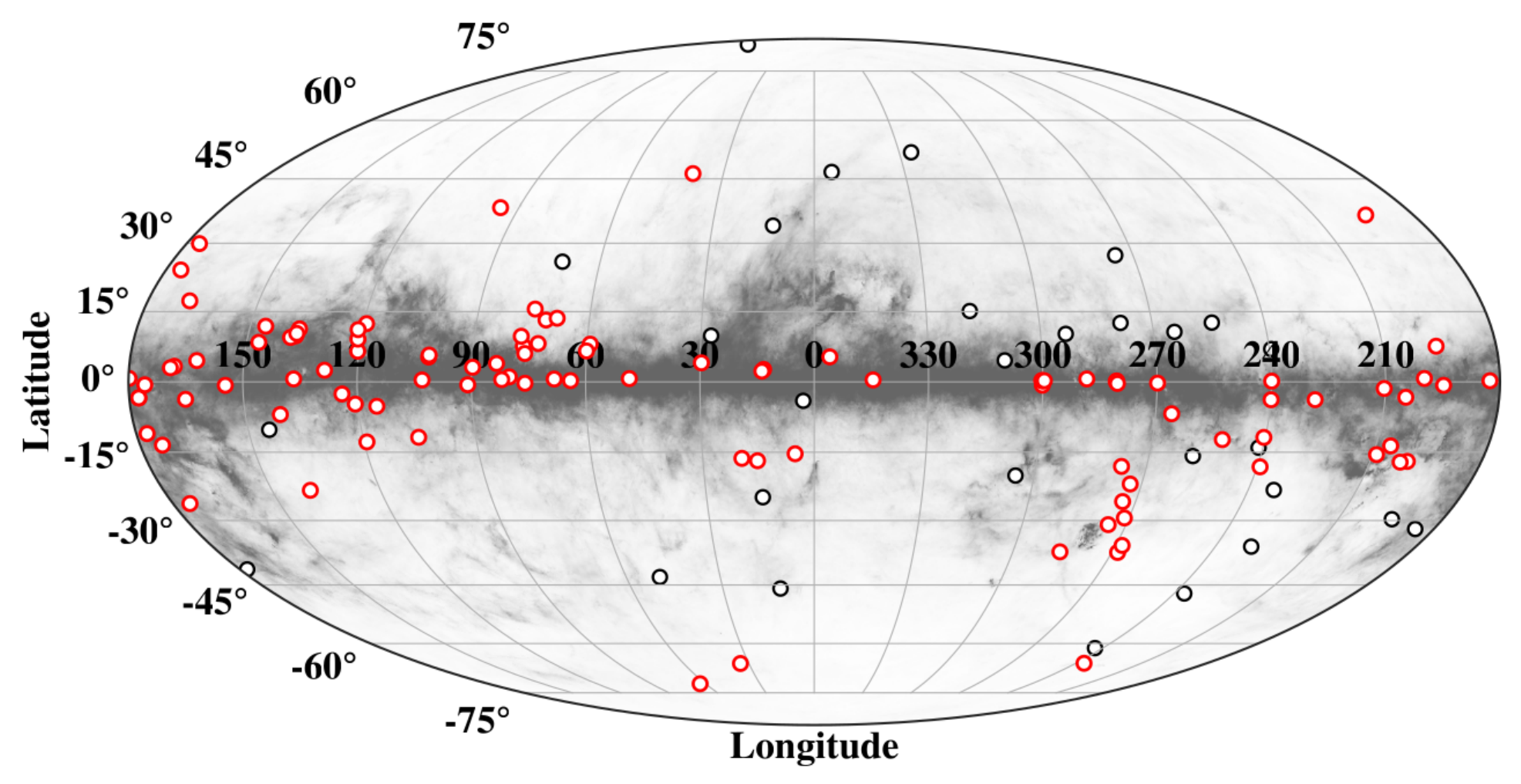}
        \end{subfigure}
        \begin{subfigure}[b]{1.\linewidth}
        \includegraphics[width=1\linewidth]{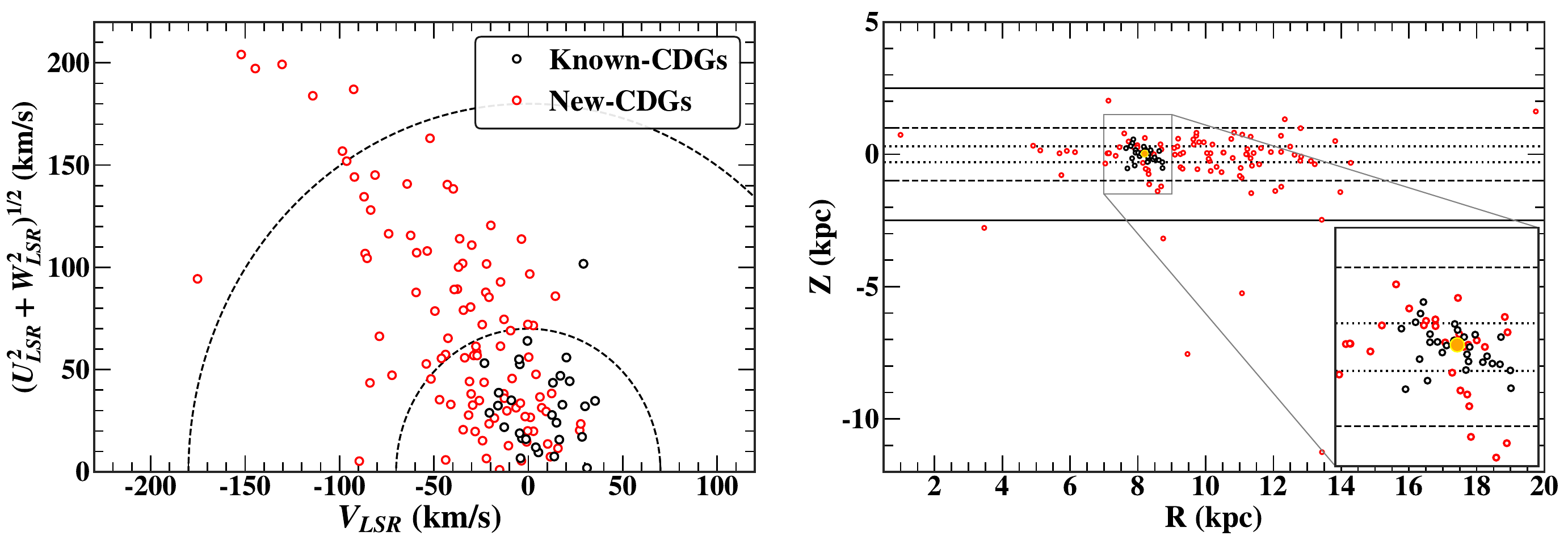}
        \end{subfigure}
        \caption{Top panel: The known-CDG (open black circles)  and the new-CDGs (open red circles) are located in the Galaxy.  The background all-sky distribution of the Galactic reddening comes from the \citet{Schlegel1998} map, as re-calibrated by \citet{Schlafly2011}. The different gray scales represent E(B - V) values from minimum (white) to maximum (grey). Bottom left panel: The known- and the new-CDGs are shown in the Toomre diagram. Dashed lines show constant values of the total space velocity, $v_{tot}$ = ($U^{2}_{LSR}$+$V^{2}_{LSR}$+$W^{2}_{LSR}$)$^{1/2}$ at 70 and 180\,km\,s$^{-1}$, demarcating thin disk and thick disk components, respectively \citep{Nissen2004, Venn2004}. Bottom right panel: Spatial distribution above/below the Galactic plane vs. the galactocentric distance. The commonly accepted scale heights of the thin disk, thick disk, and halo are indicated at $Z = \pm 0.3$~kpc (dotted lines), $Z = \pm 1$~kpc (dashed lines),  and $Z =\pm 2.5$~kpc (solid lines), respectively \citep{Sparke2007}. The Sun (filled yellow circle) is at R = 8.2~kpc and Z = 0.025~kpc.}
         \label{fig:fig2}
     \end{figure*}
  
    In bottom left panel of Fig.~\ref{fig:fig2}, we place the CDGs in the Toomre diagram. The thresholds used by \cite{Nissen2004} (also \citealt{Venn2004}) to separate the thin disk, thick disk, and halo stars are shown by dashed lines of the total space velocity, $v_{tot}$. We categorize stars with $v_{tot} < 70$\,km\,s$^{-1}$ as thin disk stars, those between $70$ and $180$\,km\,s$^{-1}$ as thick disk stars, and those with velocities greater than this as halo stars. The distribution of the CDGs in the Toomre diagram suggests that almost all of the known-CDGs (97~per cent of the known-CDGs) belong to the thin disk, except one that is likely a thick disk star. 
    
    On the other hand, the new-CDGs are found in all three components. Of the new-CDGs, 48~per cent belong to the thin disk, 44~per cent belong to the thick disk, and 8~per cent belong to the halo. Some of the new-CDGs have very large $V_{LSR}$ values with retrograde motion.

    In the bottom right panel of Fig.~\ref{fig:fig2} we show the spatial location of all the CDGs in the $R-Z$ plane. The distribution of the known-CDGs is consistent with the Toomre diagram component identifications, such that the majority of the stars are within the scale-height of the thin disk. Likewise, the new-CDGs are found in all components of the Galaxy.

    Our result for the known-CDGs is in contrast to the suggestion of \cite{Bond2019} that the known-CDGs are located at systematically larger distances from the Galactic plane than normal giants, and therefore may be thick-disk stars. Most of the known-CDG sample is from \cite{Bond2019} (27/29 stars), which is $66$~per cent of his sample. That the majority of the known-CDG sample exist mainly in the thin disk is likely due to selection bias towards brighter giants, which are in general nearby and therefore more likely to be in the thin disk. We note that $90$~per cent of the known-CDG sample are originally from the bright star catalog of \cite{Bidelman1973}. Our new-CDG sample, being from APOGEE, does not have such a bright limiting magnitude, allowing us to identify CDGs out to greater distances in the various Galactic components (Fig.~\ref{fig:fig2}).

\begin{table*}
\centering
\caption{Kinematic parameters of the known- and the new carbon-deficient giants. The full sample will be provided online and in the supplementary online-only material.}
\label{tab:table3}
\resizebox{\linewidth}{!}{%
\begin{tabular}{lrrrrrrrrrr}
\hline
\multicolumn{1}{c}{Star}      & \multicolumn{1}{c}{l}      & \multicolumn{1}{c}{b}      & \multicolumn{1}{c}{RV}               & \multicolumn{1}{c}{X}      & \multicolumn{1}{c}{Y}     & \multicolumn{1}{c}{Z}      & \multicolumn{1}{c}{R}     & \multicolumn{1}{c}{U}       & \multicolumn{1}{c}{V}                & \multicolumn{1}{c}{W}      \\
                             & \multicolumn{1}{c}{(deg)}  & \multicolumn{1}{c}{(deg)}  & \multicolumn{1}{c}{(\,km\,s$^{-1}$)} &    \multicolumn{1}{c}{(kpc)}    & \multicolumn{1}{c}{(kpc)} &    \multicolumn{1}{c}{(kpc)}    & \multicolumn{1}{c}{(kpc)} &    \multicolumn{1}{c}{(\,km\,s$^{-1}$)}     & \multicolumn{1}{c}{(\,km\,s$^{-1}$)} &   \multicolumn{1}{c}{(\,km\,s$^{-1}$)}     \\
\hline
\hline
HD 40402$^{**}$           & 232.90  & $-23.22$ & $-6.87$            & $-8.53$  & $-0.44$ & $-0.21$  & 8.55  & $+60.40$    & $-0.47$            & $-21.28$ \\
HD 124721$^{**}$          & 318.27 & $+15.14$  & $-41.44$           & $-7.63$  & $-0.51$ & $+0.23$   & 7.65  & $-43.54$  & $+13.04$            & $+1.36$   \\
37 Com$^{**}$              & 95.60   & $+85.86$  & $-14.95$           & $-8.2$   & $+0.01$  & $+0.21$   & 8.20   & $-0.84$   & $-4.11$            & $-6.61$  \\
HR 885                    & 144.44 & $-10.22$ & $+2.62$             & $-8.32$  & $+0.09$  & $+0.00$      & 8.32  & $+6.25$    & $+14.92$            & $+23.24$  \\
HR 1023$^{**}$               & 177.72 & $-41.26$ & $+5.82$             & $-8.45$  & $+0.01$  & $-0.19$  & 8.45  & $+14.74$   & $+16.41$            & $-5.58$  \\
HR 1299                   & 236.63 & $-47.05$ & $+28.37$            & $-8.31$  & $-0.16$ & $-0.18$  & 8.31  & $+40.42$   & $-4.51$            & $-33.65$ \\
HD 28932                    & 194.55 & $-29.71$ & $+17.99$            & $-8.62$  & $-0.11$ & $-0.22$  & 8.62  & $-2.19$   & $+13.75$            & $-7.21$  \\
HD 31869                 & 229.25 & $-35.94$ & $+49.16$            & $-8.69$  & $-0.57$ & $-0.52$  & 8.73  & $-31.21$  & $-16.07$           & $-8.74$  \\
HD 49960$^{**}$           & 241.16 & $-14.03$ & $+55.52$            & $-8.48$  & $-0.50$  & $-0.12$  & 8.49  & $-49.42$  & $-23.22$           & $+19.63$  \\
HD 67728$^{**}$           & 239.49 & $+7.00$      & $+12.29$            & $-8.60$   & $-0.68$ & $+0.12$   & 8.63  & $+7.86$    & $-3.35$            & $-14.41$ \\
\multicolumn{1}{c}{...}       & ...    & ...    & ...              & ...    & ...   & ...    & ...   & ...     & ...              & ...   \\

2M04403830+2554274         & 174.05 & $-13.49$ & $-34.58$           & $-10.31$ & $+0.22$  & $-0.48$  & 10.32 & $+45.52$   & $-8.58$            & $+3.63$   \\
2M12261419-6325497        & 300.12 & $-0.70$   & $-31.70$            & $-5.18$  & $-5.21$ & $-0.06$  & 7.35  & $-163.12$ & $-52.06$           & $+0.60$    \\
2M20244529+3712196         & 75.93  & $-0.29$  & $-75.93$           & $-7.17$  & $+4.11$  & $+0.00$      & 8.26  & $+62.65$   & $-78.87$           & $+21.72$  \\
2M05481354+2926231        & 179.86 & $+0.75$   & $+11.79$            & $-11.93$ & $+0.01$  & $+0.08$   & 11.93 & $-0.89$   & $-15.22$           & $-0.53$  \\
2M00344095+5804544         & 120.71 & $-4.72$  & $-66.6$            & $-10.75$ & $+4.29$  & $-0.38$  & 11.58 & $+86.71$   & $-22.53$           & $-13.8$  \\
2M12254845-6228451        & 299.98 & $+0.25$   & $-2.44$            & $-6.23$  & $-3.42$ & $+0.04$   & 7.1   & $-107.82$ & $-53.54$           & $+6.42$   \\
2M01450782+6445416         & 128.62 & $+2.48$   & $-63.56$           & $-11.06$ & $+3.59$  & $+0.23$   & 11.63 & $+54.01$   & $-33.68$           & $-14.05$ \\
2M19055092+3745351         & 68.66  & $+13.59$  & $+1.56$             & $-7.51$  & $+1.77$  & $+0.48$   & 7.73  & $+68.00$      & $-0.24$            & $-23.98$ \\
2M04222073+4844377        & 154.59 & $-0.72$  & $+7.62$             & $-11.14$ & $+1.40$   & $-0.01$  & 11.23 & $-4.83$   & $-3.64$            & $+2.51$   \\
2M16405512+6435204$^{\dagger}$        & 95.57  & $+38.16$  & $-1.04$            & $-8.22$  & $+0.17$  & $+0.16$   & 8.22  & $+27.39$   & $+9.47$             & $+11.05$  \\
\multicolumn{1}{c}{...}         & ...    & ...    & ...              & ...    & ...   & ...    & ...   & ...     & ...              & ...   \\
\hline
\end{tabular}%
}
\begin{quoting}
      { \textbf{Notes.}\\
      $^{(**)}$  Binary/ Multiple system.\\
      All the stellar designations beginning with `2M' denote the APOGEE ID of the new carbon-deficient giants.\\
      $^{(\dagger)}$ This star has been previously identified as a weak CH star by \cite{Keenan&McNeil1989}.}
\end{quoting}
\end{table*}

   \subsection{Hertzsprung–Russell Diagram} \label{sec:HRD_results}
   
    Here we attempt to estimate the evolutionary phases and masses of the CDG stars. We do this by placing them on the HRD and comparing them to model tracks.
    Luminosities were derived using the standard formula\footnote{log($L/L_{\odot}$) = $-$0.4 [$V_{0}$ $-$ $(m - M)_{0}$ $+$ $BC$ $-$ $M_{bol,\odot}$]}. Distances were taken from the catalog of \citet{Bailer-Jones2021} and V magnitudes and their errors were taken from APASS \citep{Henden2015}. In cases where V was not available in APASS, we took them from the SIMBAD database \citep{Wenger2000} or the NOMAD catalog \citep{Zacharias2004}. For some of the new-CDGs, their V magnitudes were not available in the above-mentioned catalogs, and in such cases we used the \textit{Gaia} G magnitudes and converted them to V magnitudes using the colour-colour transformations given in \citet{Riello2021}. We applied bolometric corrections (BC) to the absolute magnitudes following the relation from \citet{Alonso1999}. Reddening values were taken from the three-dimensional dust map of \citet{Green2019}. In cases where the reddening wasn't available, we took them from the \citet{Schlegel1998} map which is re-calibrated in \citet{Schlafly2011}. Five new-CDGs were of super-solar metallicity ([Fe/H] $ > +0.2$~dex) which put them outside the range of the BC relation. Hence their luminosities could not be estimated.
    In such cases, we took the luminosity value from \textit{Gaia}~DR2/DR3. Four out of five of these stars had luminosities in \textit{Gaia}. In all, we have luminosities for 101 of the new-CDGs\footnote{One star did not have its distance in the \cite{Bailer-Jones2021} catalog, and the other did not have its luminosity in \textit{Gaia}~DR2/DR3}.  We derived luminosities for the known-CDGs in the same manner. All luminosities are listed in Table~\ref{tab:table1}.

     \begin{figure*}
     \centering
    \includegraphics[scale=0.4]{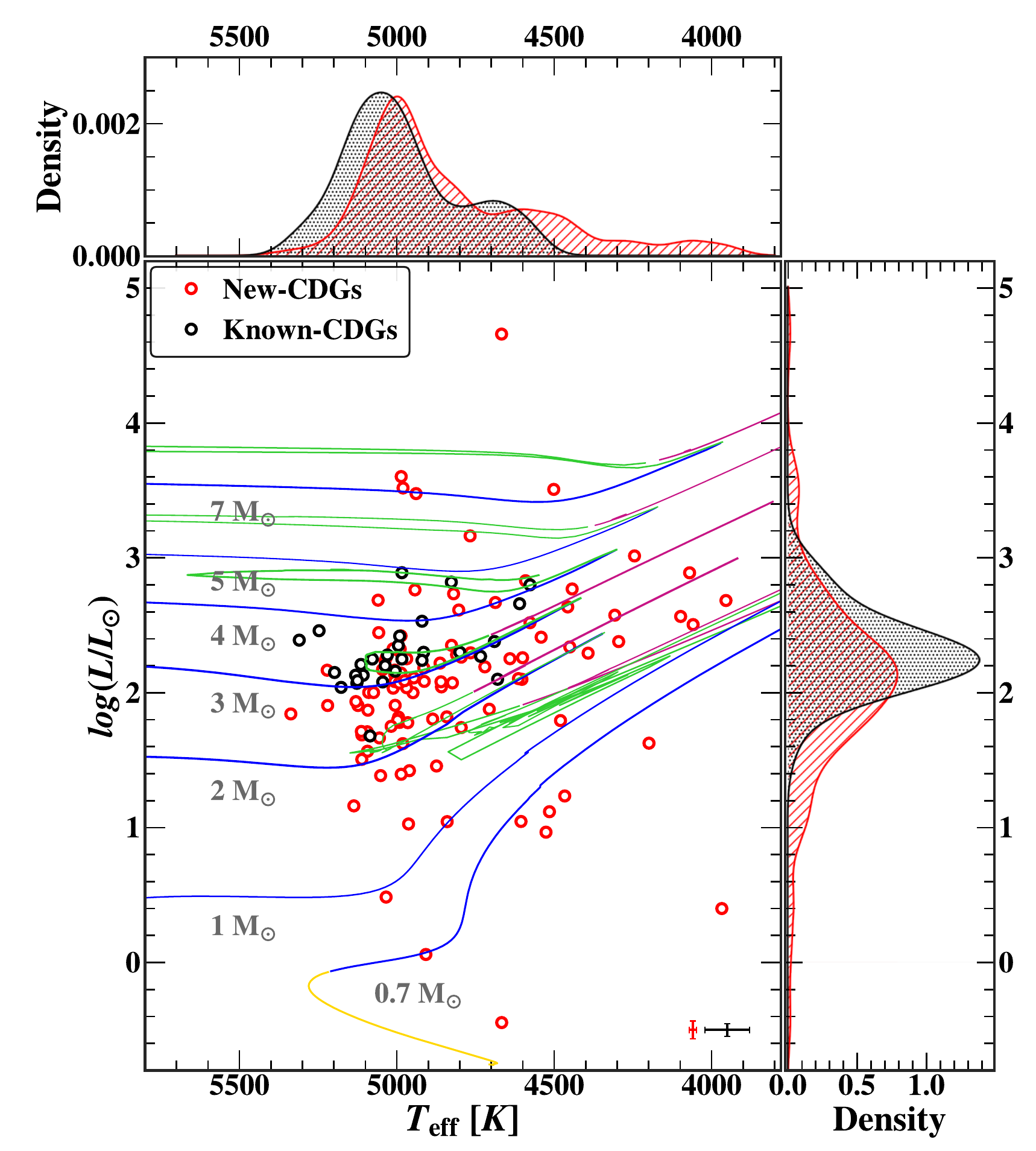}
    \caption{The known-CDGs (open black circles) and the new-CDGs (open red circles) are shown in the HRD. Model tracks of different masses with [Fe/H]$ =-0.2$~dex are superimposed for comparison. The colours of the tracks indicate particular evolutionary phases: yellow for the main-sequence phase, blue for the red giant branch phase, green for the central He-core burning phase, and magenta for the early asymptotic giant branch phase. Average error bars are shown in the bottom right corner. Kernel density histograms are included to show the temperature (top panel) and luminosity distributions (right panel) of the CDGs, with the colour scheme following the sample.}
     \label{fig:fig3}
    \end{figure*}

    Effective temperatures ($T_{\rm eff}$) are taken from the APOGEE catalog for the new-CDGs and from the literature \citep{Adamczak2013, Palacios2016} for the known-CDGs.

    We show both the new- and known-CDGs in the HRD in Fig.~\ref{fig:fig3}. The CDGs are typically slightly metal-poor, with an average [Fe/H$] =-0.2$~dex. Hence we plot evolutionary tracks with [Fe/H$] =-0.2$~dex for comparison. The models have a range of masses, as indicated. The evolutionary tracks were taken from the MIST database\footnote{\url{http://waps.cfa.harvard.edu/MIST/}}. The colours of the tracks indicate the various evolutionary phases.

    It can be seen that the new-CDGs have a wider range in luminosity and $T_{\rm eff}$ compared to the known-CDGs. This may be an artefact resulting from the small sample size of the known-CDGs, since the peaks of the distributions in $\log(L/L_{\odot})$ and $T_{\rm eff}$ are similar (right and top panels of Fig.~\ref{fig:fig3}). Roughly 75~per cent of the stars in the new-CDG sample overlap with the known-CDG sample luminosity range.

    Looking at the distribution of the luminosities in the right panel of Fig.~\ref{fig:fig3}, it seems the known-CDGs may have slightly higher luminosity than the new-CDGs, on average. Taking a straight average, the known-CDGs have $\log(L/L_{\odot}) = 2.30 \pm 0.05$~dex and the new-CDGs $\log(L/L_{\odot}) = 2.07 \pm 0.07$~dex, where the uncertainties are the errors on the means. However, looking at the peaks of the distributions (modes), we see that they are close ($\log(L/L_{\odot}) = 2.23$ vs $2.13$~dex). The modes differ by $0.1$~dex, which is within the 1$\sigma$ uncertainties mentioned above. Hence we conclude that the luminosity peaks are largely consistent with each other, which is also evident in the luminosity histogram of Fig.~\ref{fig:fig3}.

    The top panel of Fig.~\ref{fig:fig3} shows that the known-CDGs have slightly higher temperatures on average than new-CDGs. The modes differ by $\approx 50$~K. This magnitude of difference could be explained by systematics between the studies, for example, the new-CDG temperatures are based on IR spectra and the known-CDGs on optical spectra. Apart from a small extended tail in the new-CDGs, we conclude that the $T_{\rm eff}$ distributions are also similar between the known- and new-CDGs.

    In summary, in the HRD, our sample of new-CDGs is quite similar to the previously known CDGs.

    With most CDGs having $T_{\rm eff} \approx 5000$~K,  the majority are too warm to be stars on the RGB ($T_{\rm eff} \lesssim 4500$~K; Fig.~\ref{fig:fig3}).  These temperatures are much more consistent with them being in the subgiant branch (SGB) phase or core helium-burning phase (red clump; RC). This is in good agreement with \citet{Palacios2012}. The cooler stars (second peak in $T_{\rm eff}$ distribution, around $4700$~K) are also consistent with either the SGB or the RC, but also the early-asymptotic giant branch (EAGB). The rarer cooler CDGs may also be consistent with being RGB stars. It can be seen that there is substantial degeneracy in terms of determining the evolutionary phase(s) of the CDGs. Asteroseismology of these objects would likely help in determining their evolutionary state (e.g.~\citealt{Bedding2011, Mosser2014}).

     In terms of mass, we can see that the $3~\rm{M}_{\odot}$ track passes through the main peak of the CDG distribution in the HRD, and the peak is bracketed by the $2$ and $4~\rm{M}_{\odot}$ models. Thus we conclude that the majority of the CDGs are intermediate-mass stars. This is consistent with earlier studies \citep{Palacios2012, Palacios2016,Bond2019}. 
     
     The extended distribution of the new-CDGs in the HRD may indicate that CDGs can also sometimes have lower and higher masses. We inspected these stars closely, and found that the stars with outlying luminosities are  located in the Galactic plane. They have very high reddening values (E(B-V) = 2 -- 11~mag; Table~\ref{tab:table1}). Although the formal error bars (which are propagated to the luminosity uncertainties) on the reddening values are small (\citealt{Green2019,Schlegel1998,Schlafly2011}), we believe these must be underestimated for regions of such high dust content. Thus we treat the highest and lowest luminosity stars with caution. The upshot is that the tails of mass distribution inferred from the HRD (Fig.~\ref{fig:fig3}) may not be real.

     The (inferred) low-mass CDGs ($M < 2.0~\rm{M}_{\odot}$) have not been seen in previous studies \citep{Palacios2012, Palacios2016,Bond2019} \footnote{During our study, \cite{Holanda2023} discovered one low-mass CDG, HD~16424. It has a mass of~$1.61 \pm 0.26~\rm{M}_{\odot}$, based on its location in the HRD and also a Bayesian estimation method.}.

\subsection{Abundances: Carbon, Nitrogen, and Oxygen} \label{sec:cno_results}

\begin{figure*}
    \centering
    \includegraphics[width=\linewidth]{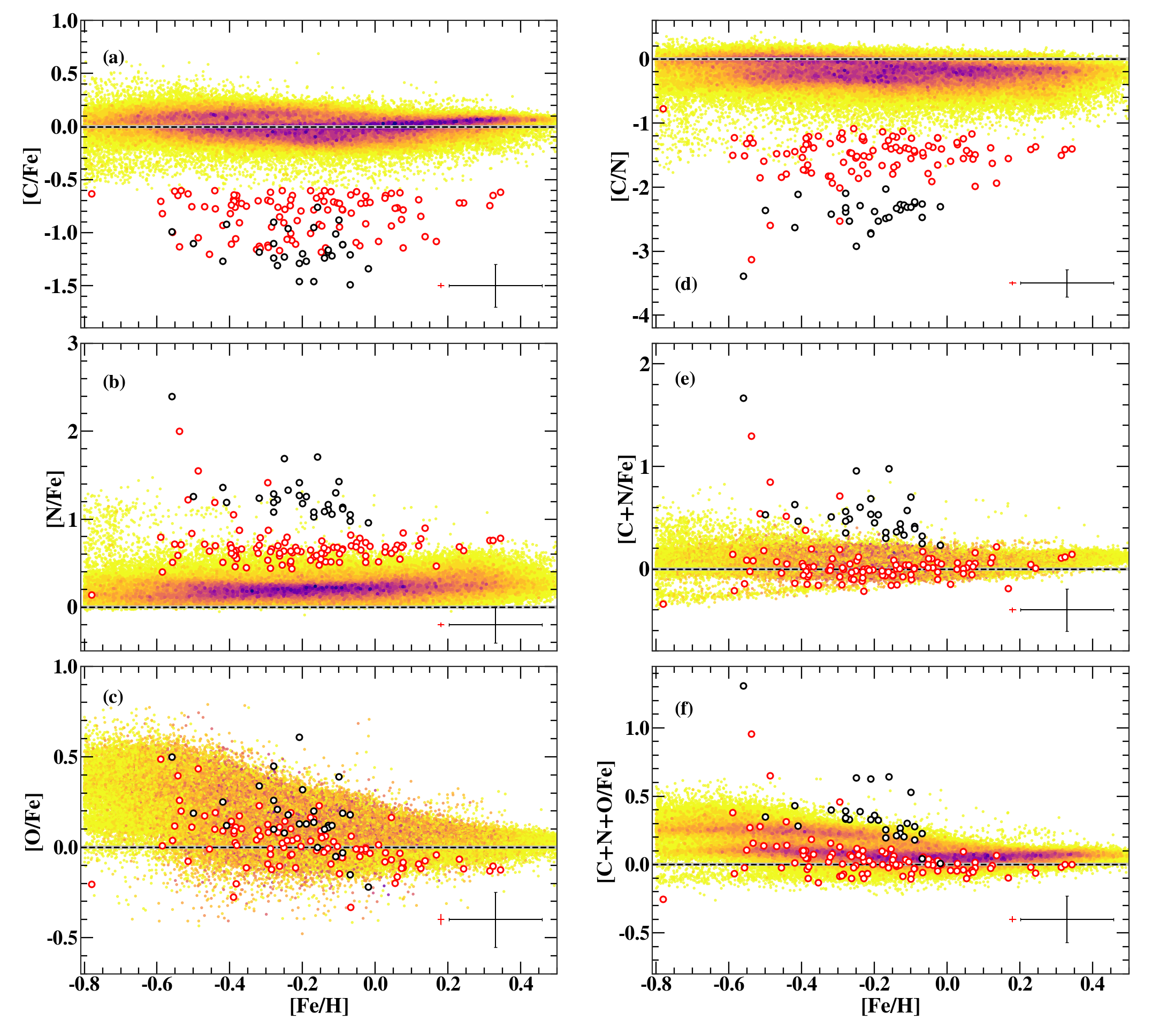}
    \caption{C, N, and O abundance ratios against [Fe/H] for the known-CDGs (open black circles) and the new-CDGs (open red circles). The selected sample giants from the APOGEE~DR17 survey (315,789 stars; see Section~\ref{sec:data}) form the background in a colour scale that represents the number density of stars; darker colours indicate higher density. We apply the average offset of $+0.02$, $-0.15$, $-0.12$, and $-0.03$~dex to the [C/Fe], [N/Fe], [O/Fe], and [Fe/H] values of known-CDGs respectively, to  account for the systematic difference between the abundances from IR and optical spectra (\citealt{Jonsson2020}; see text for details). Average error bars are shown on the lower right of each subplot. Solar abundances are from \citet{Grevesse2007}.}
    \label{fig:fig4}
    \end{figure*}

    In Fig.~\ref{fig:fig4} we plot the C, N, and O abundances, the [C+N/Fe] and [C+N+O/Fe] sums, and the [C/N] ratio against [Fe/H] for a background sample of APOGEE giants, along with the known- and new-CDG stars. The abundances of the background giants and the new-CDGs are taken from the APOGEE catalog (DR17; \citealt{Abdurrouf2022}; \citealt{Garcia2016}). All stars have good quality abundances (\texttt{X\_FE\_FLAG=0}). As mentioned in Section~\ref{sec:CDG_identification}, the C, N, and O abundances of the known-CDGs are from high-resolution optical spectra \citep{Adamczak2013, Palacios2016}. In order to compare the abundances of the known-CDGs with that of our new-CDG stars and the background sample of giants, they need to be on the same scale. There are known offsets between the abundances of C, N, O, and Fe of giants (log $g < 3.5$~dex) derived from APOGEE IR spectra and the abundances derived from high-resolution optical studies. For the [C/Fe], [N/Fe], [O/Fe], and [Fe/H] abundance ratios determined from high-resolution optical spectra to be on the same abundance scale as APOGEE, offsets of $+0.02$~dex, $-0.15$~dex, $-0.12$~dex, and $-0.03$~dex must be applied, respectively \citep{Jonsson2020}. The above-mentioned offsets are applied to the abundance ratios of the known-CDGs and are shown in  Fig.~\ref{fig:fig4} (the offsets are also applied in  Fig.~\ref{fig:fig1}).

    True to their definition, the average carbon abundances of both the new- and known-CDGs are significantly lower, by  $-0.8$~dex and $-1.2$~dex, over the average [C/Fe] of the background sample of giants\footnote{We note that the average [C/Fe] abundance of the new-CDGs would decrease if the poor-quality data CDGs of Section~\ref{sec:CDG_identification} were found to be reasonably close to their real abundances.}. The nitrogen abundances of the new-CDGs are enhanced by an average of $+0.5$~dex over the average [N/Fe] of the background sample. The known-CDGs have extremely high nitrogen abundances, their average nitrogen abundances are $+1.1$~dex higher than the background sample. We see that there are 6 new-CDGs that have similar nitrogen abundances to those of the known-CDGs. 
    
    It can be seen that the new- and the known-CDGs have very low [C/N] ratios. This stems from the extreme carbon depletion and nitrogen enhancement. The average [C/N] offset is very large for both the new- and the known-CDGs, at $+1.3$~dex and $+2.2$~dex respectively. In contrast, the [O/Fe] values of both the new- and the known-CDGs generally track the abundances of the background sample. The majority of the new-CDGs are close to scaled solar, with an average oxygen abundance of [O/Fe$] = 0.0 \pm 0.1$~dex. However, the known-CDGs have slightly higher average oxygen of [O/Fe$] = 0.2\pm 0.1$~dex, and two stars appear to have even higher [O/Fe].

    Most of the new-CDGs are scaled solar in terms of [C+N/Fe]. Six stars buck this trend, having [C+N/Fe$] > 0.3$~dex. These stars also have particularly high nitrogen abundances, being as enhanced as the known-CDGs. These trends are the same in the [C+N+O] diagram. 

\begin{figure*}
    \centering
    \includegraphics[width=0.8\linewidth]{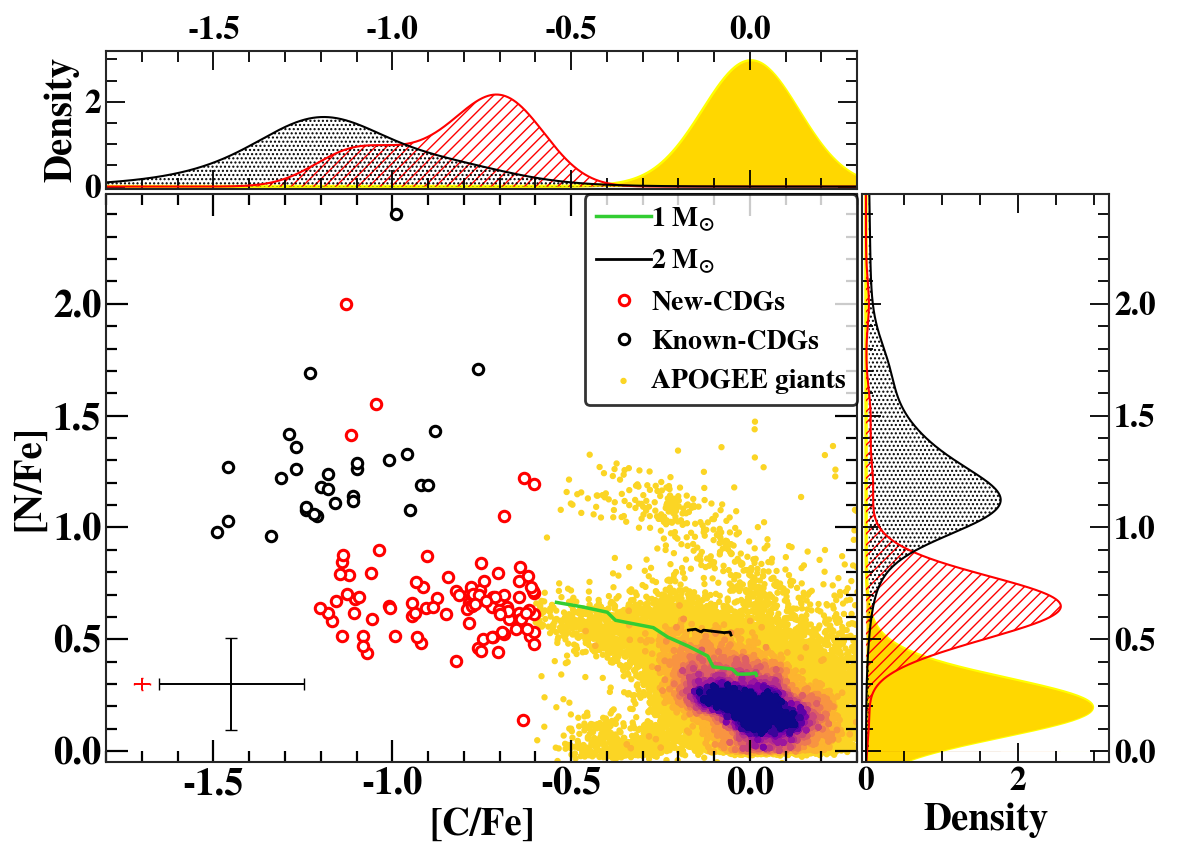}
    \caption{The known-CDGs (open black circles) and the new-CDGs (open red circles) in the [C/Fe] versus [N/Fe] plane. The APOGEE giants form the background in a colour scale that represents the number density of stars; darker colours indicate higher density. The observed [C/Fe] and [N/Fe] values are compared with the predicted [C/Fe] and [N/Fe] ratios from the standard models for initial masses 1~M$_{\odot}$ (green solid line) and 2~M$_{\odot}$ (black solid line) for the sample metallicity range for giants close to the RGB tip. Average error bars are shown on the bottom left end. Solar abundances are from \citet{Grevesse2007}.}
    \label{fig:fig5}
    \end{figure*}

    In Fig.~\ref{fig:fig5} we plot [C/Fe] vs [N/Fe]. Generally, we see that low C is associated with high N, as expected. Looking more broadly, there appears to be three `arms' in the distribution reaching towards low C: a lower arm ([N/Fe$] \lessapprox +0.3$~dex), a middle arm ($+0.3 \lessapprox [$N/Fe$] \lessapprox +0.9$~dex)  and an upper arm ([N/Fe$ \gtrapprox +0.9$). Almost all of our new-CDGs are in the middle arm, whereas the known-CDGs are primarily in the upper arm. The CDGs appear to be at the extreme extensions (very low C) of these two arms. We note that \cite{Schiavon2017} found two arms in their sample, although their sample is of lower metallicity than ours ([Fe/H$] < -0.6$~dex). In Fig.~\ref{fig:fig5} we also plot model tracks showing the carbon and nitrogen abundances at the tip of RGB for giants of two different masses, 1 and 2~M$_{\odot}$ for a range of metallicity values that cover our sample. It can be seen that the 1~M$_{\odot}$ models track the middle arm fairly well. These models include CN(O) cycling of the RGB envelope via thermohaline mixing (\citealt{Dotter2016,Choi2016}). However, the models do not reach the CDGs, which are even more C-depleted. On the other hand, the known-CDGs have N abundances well above thermohaline mixing predictions. 

    In summary, the majority of the \emph{new}-CDGs show a combination of carbon under-abundance, nitrogen over-abundance, and solar [C+N+O/Fe]. This chemical pattern is a signature of CN(O) cycling. The fact that O is not depleted suggests that only CN cycling has operated on this material (i.e. the ON cycle was not operating). 
    
    In contrast, the known-CDGs have very high [N/Fe] and [C+N+O/Fe] ratios enhanced by an average of $+0.3$~dex over the background. Enhanced [C+N+O/Fe] is a signature of CNO elements being added to the envelopes of these stars, likely through the dredge-up of He-burning products. The main He-burning product is C. That their C abundances are so low (and their N abundances so high) is an indication that there has been very strong hydrogen burning in addition to this dredge-up. This is further supported by the observed low $\rm^{12}C/^{13}C$ ratios (=~3 to 10; \citealt{Adamczak2013, Palacios2016}) and very low [C/N] ratios (Fig.~\ref{fig:fig4}d), both indicating equilibrium (or near-equilibrium) CN(O) cycling.

    The analysis above is based on our high-quality APOGEE sample. As detailed in Section~\ref{sec:CDG_identification} we identified that there is likely a substantial number of CDGs with [C/Fe]~$\lessapprox -1.2$~dex which were missed due to poor-quality APOGEE results. These stars are likely members of the same chemical group as the known-CDGs, that is they are of extreme chemical composition. If confirmed with better data, this would mean that stars with extreme composition are relatively common, which would have implications for proposed formation scenarios.

\section{Discussion and conclusion}\label{sec:discussion}

    The origin and nature of the carbon-deficient giants has been a mystery for many decades (e.g. \citealt{Bidelman1951}; \citealt{Sneden1978}; \citealt{Cottrell1978}; \citealt{Parthasarathy1980}; \citealt{Palacios2012}; \citealt{Adamczak2013}; \citealt{Palacios2016}; \citealt{Bond2019}). The previous sample in the literature has been small, with only  44 known CDGs (\citealt{Bidelman1951}; \citealt{Bidelman1973}; \citealt{Bond2019}). The first dedicated spectroscopic survey for CDGs was by \citet{Bidelman1973}, followed by \cite{Bond2019} recently. Our study, based on the high-resolution APOGEE spectroscopic survey, has increased the number of CDGs by more than a factor of 3, adding 103 newly-identified carbon-deficient stars.

    Our analysis of the CDG sample revealed a few key results:
    \setlist[enumerate]{label={\arabic*.}}
    \begin{enumerate}
        \item There is a continuous distribution in [C/Fe] deficiency, which can be seen as a tail in the [C/Fe] distribution (see right panel of Fig.~\ref{fig:fig1}). This answers the question of  \citet{Adamczak2013}, ``... concern[ing] the apparent clean distinction between normal and wGb giants -- do stars with intermediate C abundances exist in detectable numbers?". We also find that the new-CDGs represent only $0.03$~per cent of the APOGEE giants and hence are very rare.
        \item The kinematics of the sample revealed that CDGs are found in all components of the Galaxy (Section~\ref{sec:kinematics}). This is in contrast with the view in the literature that CDGs may mainly be thick-disk objects given their large distances from the Galactic plane \citep{Bond2019}.
        \item By placing the sample in the HRD (Fig.~\ref{fig:fig3}), we were able to make estimates of mass and evolutionary phase. We found that the CDGs appear to be primarily in the mass range $2-4~\rm{M}_{\odot}$ and are in the RC or SGB phase. The extended distribution of the new-CDGs in the HRD may indicate that CDGs can also sometimes have masses less than  $2~\rm{M}_{\odot}$. This is at odds with the understanding in majority of the literature, where CDGs were considered only to be of intermediate mass ($M = 2.5 - 5.0~\rm M_{\odot}$). Due to the degeneracy of the stellar tracks in the HRD, it is difficult to isolate the exact evolutionary phase(s) of the CDGs.
        \item The CNO abundances (Fig.~\ref{fig:fig4}) and the [N/Fe] vs [C/Fe] plot (Fig.~\ref{fig:fig5}) revealed that there are two groups of CDGs. The first shows the signature of CN(O) cycling of initially scaled-solar material, whilst the second shows the signature of dredge-up of He-burning products which then also underwent CN(O) cycling to reduce C and increase N. 
        We find that O is likely unchanged in most of the CDGs, indicating that the material has been processed through the CN cycle, and not the ON cycle, although there are a few known-CDGs with enhanced O.
        \item Contrary to previous studies, we find that the CDGs have a normal distribution of metallicities (Fig.~\ref{fig:fig1}).
    \end{enumerate}

    The persistent question related to the CDGs has been why some giants have extremely low C abundances compared to theoretical predictions and most of the observed giants do not. To this day, there is no consensus in the literature on this.
    
    Using updated observations and models, \citet{Palacios2016} showed that rotation-induced mixing during the MS of intermediate-mass stars (a pollution mechanism suggested by \citealt{Adamczak2013}) can decrease C and increase N, but nowhere near to the extent seen in the CDGs. We note that, observationally, only a few known-CDGs were actually found to be high-rotation giants (\citealt{Palacios2012,Adamczak2013,Palacios2016}). \citet{Palacios2016} speculated that the chemical patterns may have been imprinted earlier, at birth or during the pre-main sequence (PMS). However this theory had weaknesses, for example, there are no known MS stars with the same chemical pattern as the CDGs that could be progenitors. \citet{Palacios2016} concluded that they could find no clear explanation for the CDGs -- leaving the ``puzzle largely unsolved".

    More recently, \cite{Bond2019} suggested that CDGs may be the result of mass transfers or merger events (also see \citealt{Metzger2021,Matsumoto2022}). This was based on three pieces of evidence: (i) the systematically high distances of the CDGs from the Galactic plane as compared to normal red giants lying in the same location in the colour–magnitude diagram (CMD), (ii) their apparent high masses ($2.0 - 4.5~\rm{M}_{\odot}$), which might imply that they are binaries or binary products \citep{Izzard2018}, and (iii) some of the CDGs having high rotation rates.
    In the current study we confirm that the CDGs are primarily intermediate-mass stars, although they also cover a wide range of masses (Section~\ref{sec:HRD_results}), which may be difficult to reconcile with the merger scenario. Importantly, we also show that there is no preference for Galactic location amongst the CDGs (Table~\ref{tab:table3}; Fig.~\ref{fig:fig2}), which counters one of the key pieces of evidence for the merger scenario.
    
    To explore the merger scenario further we checked for binarity in the literature \citep{Palacios2012, Adamczak2013, Palacios2016}, and various catalogs (The Washington Visual Double Star catalog \citep{Mason2001}, Catalog of Components of Double and Multiple stars \citep{Dommanget2002}, an extensive catalog constructed from \textit{Gaia}~eDR3 binary stars within $\approx 1$~kpc of the Sun\footnote{We note that \textit{Gaia} catalogue is of spatially resolved binaries only.} \citep{El-Badry2021}, and the Binary Star Database \citep{Kovaleva2015}). We found 12 known-CDGs either in multiple systems or in a binary system. This is about $41$~per cent of the 29 known-CDGs with carbon abundances, which is broadly consistent with the expected binary fraction in this mass range ($\sim 48$~per cent; \citealt{Parker2014}). This suggests that binarity is not a key factor in the formation of the CDGs, and is another piece of evidence against the merger scenario. We do not have information on the binary status of our sample of new-CDGs.

    Taking our results into account, we explore the scenarios that might be considered to account for the chemical peculiarities of the CDGs. Having [C+N+O/Fe]~$> 0.0$ is a signature of He-burning products being mixed up to the envelope, while [C+N+O/Fe]~$ =0.0$ is a signature of CN(O) cycling only. Given that our CDG sample has scaled-solar oxygen (Fig.~\ref{fig:fig4}; Table~\ref{tab:table1}) and [C+N+O/Fe]~$=0.0$, it appears that the ON cycle was not activated in these stars. 
    
    The known-CDGs have much higher N abundances than our sample of CDGs. These stars must have had CNO elements added to their envelopes, likely through dredge-up of He-burning products. This is conveyed by their overabundances of [C+N+O/Fe] (Fig.~\ref{fig:fig4}). However, the fact that the N is so enhanced and C is so depleted indicates that there has been very strong hydrogen burning on top of this. This is supported by the low observed $\rm^{12}C/^{13}C$ ratios (=~3 to 10; \citealt{Adamczak2013, Palacios2016}) and very low [C/N] (see Fig.~\ref{fig:fig4}d), both indicating equilibrium (or near equilibrium) CN(O) cycling.

    It is quite evident from this study that we cannot speculate further without more data. Asteroseismic data would be very useful for mass determinations ($\nu_{max}$, $\Delta \nu$), and to lift the degeneracy of the evolutionary status of the CDGs between subgiant branch, red giant branch, and core helium-burning phases ($\Delta$P; \citealt{Bedding2011}). It is also important to conduct detailed abundance studies of all the CDGs including the additional CDGs from APOGEE sample with WARN Flags. Further, a dedicated binary survey would be required to be certain of the true binary fraction. In order to unravel the mystery of the origin of the CDGs, it is important to determine whether C-deficiency correlates with rotation, mass, binarity, evolutionary phase, and carbon isotopic ratios (to name a few parameters). Presently, most of the CDGs lack these parameters. In our following study, we will focus on their evolutionary phase and mass determinations using asteroseismology.

\section*{Acknowledgements}
    The authors thank the referee for comments that have helped improve the overall quality of the paper and also clarify the work presented here. This study is supported by the National Natural Science Foundation of China under grant Nos. 11988101, 11890694, and 11873052 and National Key R\&D Program of China No.2019YFA0405500. S.M acknowledges the support of the CAS-TWAS Presidents fellowship for International Doctoral Students. S.M. thanks Sarah A. Bird for helpful conversations. Y.B.K. thanks M. Parthasarathy for introducing this rare group of stars. We thank Evgenii Neumerzhitckii for the use of his plotting routine. This work has made use of data from the European Space Agency (ESA) mission \textit{Gaia} (\url{https://www.cosmos.esa.int/gaia}). We made use of the SIMBAD database and the VizieR catalog access tool, CDS, Strasbourg, France. We thank Evgenii Neumerzhitckii for the use of his plotting routine. S.W.C. acknowledges federal funding from the Australian Research Council through a Future Fellowship (FT160100046) and Discovery Projects (DP190102431 \& DP210101299). Parts of this research was supported by the Australian Research Council Centre of Excellence for All Sky Astrophysics in 3 Dimensions (ASTRO 3D), through project number CE170100013. This research was supported by use of the Nectar Research Cloud, a collaborative Australian research platform supported by the National Collaborative Research Infrastructure Strategy (NCRIS).

\section*{Data Availability}

    The data in this article (Table~\ref{tab:table1}, Table~\ref{tab:table2} and Table~\ref{tab:table3}) will be available in the online-only supplementary material and an online catalog.

\bsp	
\label{lastpage}
\end{document}